\documentclass[11pt,a4paper]{article}
\usepackage[final]{arr}
\usepackage{times}
\usepackage{latexsym}
\usepackage{soul}
\usepackage{booktabs}
\usepackage{amsmath,caption}
\usepackage{colortbl}
\usepackage{arydshln}
\usepackage{multirow}
\usepackage{multicol}
\usepackage{hyperref}

\usepackage{xspace}
\usepackage{adjustbox}
\usepackage{subcaption}
\usepackage{longtable}
\usepackage{makecell}

\usepackage{microtype}
\usepackage{graphicx}
\usepackage{multirow}

\newcommand\wtq{{WikiTQ}\xspace}
\newcommand\tapex{\textsc{TaPEx}\xspace}
\newcommand\ours{{TaCube}\xspace}
\newcommand\uniskg{\textsc{UnifiedSKG}\xspace}

\title{TaCube: Pre-computing Data Cubes for Answering Numerical-Reasoning Questions over Tabular Data}

\author{Fan Zhou\textsuperscript{1}$\thanks{$^{*}$Work done during internship of Fan and Mengkang at Microsoft Research Asia.}$, ~Mengkang Hu\textsuperscript{2},~Haoyu Dong\textsuperscript{3}$\thanks{$^{\dagger}$Corresponding author.}$ , ~Zhoujun Cheng\textsuperscript{1},~Shi Han\textsuperscript{3}, ~Dongmei Zhang\textsuperscript{3}
     \\
     ~\textsuperscript{1} Shanghai Jiao Tong University, ~\textsuperscript{2}Harbin Institute of Technology
     ~\textsuperscript{3}Microsoft Research \\
\{zhoufan98,blankcheng\}@sjtu.edu.cn,
1190200505@stu.hit.edu.cn,\\
\{hadong, shihan, dongmeiz\}@microsoft.com}

\date{}

\begin{document}
\maketitle

\begin{abstract}
Existing auto-regressive pre-trained language models (PLMs) like T5 and BART, have been well applied to table question answering by \uniskg and \tapex, respectively, and demonstrated state-of-the-art results on multiple benchmarks. However, auto-regressive PLMs are challenged by recent emerging numerical reasoning datasets, such as TAT-QA, due to the error-prone implicit calculation. In this paper, we present \ours,  to pre-compute aggregation/arithmetic results for the table in advance, so that they are handy and readily available for PLMs to answer numerical reasoning questions. 
\ours systematically and comprehensively covers a collection of computational operations over table segments.
By simply concatenating \ours to the input sequence of PLMs, it shows significant experimental effectiveness. \ours promotes the F1 score from 49.6\% to 66.2\% on TAT-QA and achieves new state-of-the-art results on \wtq (59.6\% denotation accuracy). 
\ours’s improvements on numerical reasoning cases are even more notable: on TAT-QA, \ours promotes the exact match accuracy of BART-large by 39.6\% on $\mathtt{sum}$, 52.5\% on $\mathtt{average}$, 36.6\% on $\mathtt{substraction}$, and 22.2\% on $\mathtt{division}$.
We believe that \ours is a general and portable pre-computation solution that can be potentially integrated to various numerical reasoning frameworks. Data and code will be available at \url{https://github.com/koalazf99/tacube}. 



\end{abstract}

\section{Introduction} \label{sec:intro}

\begin{figure}[t]
    \begin{center}
    \includegraphics[width=0.45\textwidth]{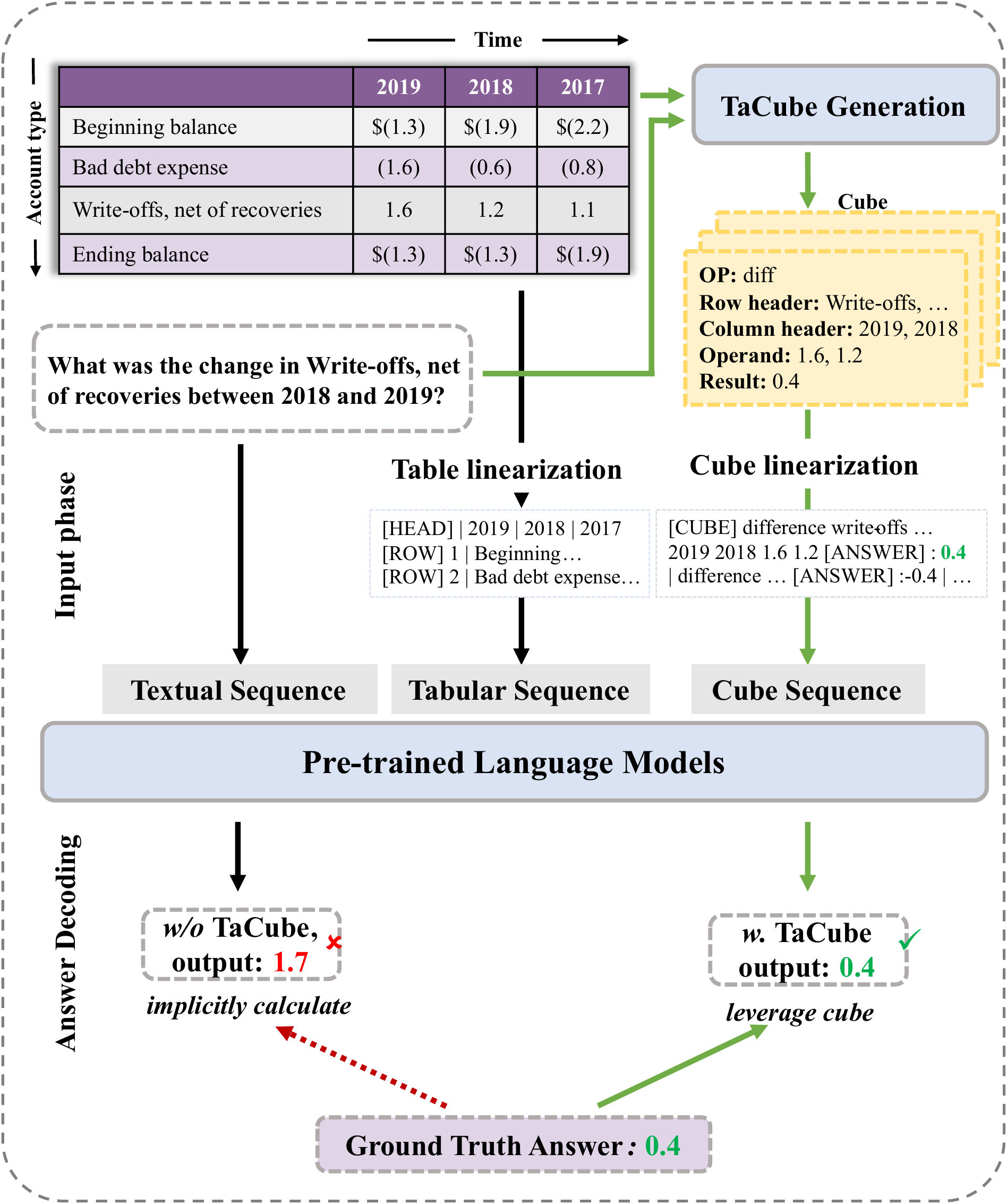}
    \end{center}
    \vspace{-0.3cm}
\caption{Augmenting auto-regressive PLMs with \ours. By simply concatenating \ours to the input sequence, \ours significantly mitigates the calculation challenge in numerical reasoning over tables.}
\label{fig:intro}
\vspace{-0.32cm}
\end{figure}

There are a flurry of recent works on table-text joint reasoning, e.g., answering questions over tables~\cite{yu2018spider, pasupat2015compositional,zhong2017seq2sql}. Meanwhile, pre-trained language models (PLMs), which have demonstrated great success on various natural language (NL) tasks, are also well applied to table-text joint reasoning and show great effectiveness~\cite{yin2020tabert, herzig2020tapas, cheng2021fortap,liu2021tapex,xie2022unifiedskg,dong2022table}. 

Recently, numerical reasoning (NR) over tabular data has raised increasing attention and a bunch of table QA datasets targeting numerical reasoning have been collected~\cite{chen2021finqa, zhu2021tat, zhao2022multihiertt, cheng2021hitab}, promoting NR to be a hot challenge in table QA.
However, faced with flexible calculations such as addition, comparison, and aggregation over semi-structured context, PLMs encounter great obstacles ~\cite{liu2021tapex, zhu2021tat}, especially when calculation skills have not been fully exploited during large-scale pre-training on NL corporal.


Existing approaches to mitigate this gap can be mainly concluded into two families. One is logical-form-based method, which formulates table QA as a semantic parsing task. The method first generates the logical form and then applies a post execution on the logical form to produce the final answer. The answer generation process is much more trackable and explainable than directly decoding answers, but it also has deficiencies. (1) Human annotations of logical forms are expensive and error-prone~\cite{herzig2020tapas, cheng2021fortap}. 
(2) There lacks a dominant formulation of logical forms for table reasoning, and thus logical form formulations and model designs are easy to be task-specific and even dataset-specific: ~\cite{pasupat2015compositional,liang2018mapo, guo2019towards, gao2019hybrid, cheng2021hitab} design their own logical forms and semantic parsers; ~\cite{herzig2020tapas, zhu2021tat} simplify the logical form design and use specific classification heads for cell selection, operator prediction, and unit prediction, to produce the answer. 
(3) Due to the representation limitation of logical forms, it lacks sufficient flexibility to generate free-style answers or answers matching special conditions, e.g., ``2 million" that augments a numeric value "2" with its textual scale "million" and "2022" that is a part of a cell string "2022/05/20".



The other popular way is directly generating the answer using auto-regressive PLMs. \uniskg~\cite{xie2022unifiedskg} leverages T5 ~\cite{raffel2020t5} and achieves promising and even state-of-the-art performance on a series of datasets, showing the power of directly fine-tuning large LMs on table-text joint reasoning. \tapex~\cite{liu2021tapex}, which is based on BART~\cite{lewis2020bart} and pre-trained by learning a neural SQL executor, surprisingly outperforms prior works by a large margin. \tapex provides a new way of using clean and synthetic data for efficient pre-training and shows its promising capacity on several numerical reasoning types, e.g., $\mathtt{count}$ and $\mathtt{comparison}$. 
However, when it comes to more complex numerical calculations involving $\mathtt{sum}$, $\mathtt{substraction}$, or $\mathtt{division}$, the accuracy falls to the bottom, showing that implicit numerical calculation is error-prone. 
To make things worse, when auto-regressive PLMs produce a wrong number, e.g., "1.7" in Figure~\ref{fig:intro}, it’s hard to uncover where the number comes from and improve the model accordingly, because all calculations are implicitly done in transformers.


To address the above shortcomings, we propose to pre-compute aggregation/arithmetic results for the target table in advance, so that they are handy and readily available for answering numerical reasoning questions. This idea is inspired by an important data cube~\cite{jiawei2016data} concept in data mining field to facilitate the online analytical processing of multi-dimensional data, so we name our pre-computed aggregation/arithmetic results as \ours. \ours systematically covers a collection of computational operations over table segments in a comprehensive way. \ours can not only be fully materialized, but also be partially materialized for efficiently application to existing 
models. We propose rule-based and neural-based methods to produce efficient \ours while maintain high coverage on ground truth numerical reasoning. In this paper, we applied \ours to auto-regressive PLMs, for their flexibility of decoding answers by leveraging both the original table and \ours (as a fast access) to avoid most error-prone implicit calculations. We believe that \ours is a general pre-computation solution that can be helpful to various numerical reasoning frameworks.  

Our experiments show that, by directly augmenting the original table with sequenced \ours: (1) on TAT-QA, \ours significantly improves BART-large by 18.3\%  in F1 score and \tapex by 16.6\% in F1 score, and outperforms the logical-form-based state-of-the-art method TagOP by 3.5\% in F1 score; (2) on \wtq, \ours also achieves new state-of-the-art denotation accuracy of 59.6\% (+2.1\%). In addition to the overall improvements, we analyze \ours's EM improvements by different calculation operators on TAT-QA based on BART-large: $\mathtt{sum}$ increased by 39.6\%, $\mathtt{average}$ increased by 52.5\%, $\mathtt{substraction}$ increased by 36.6\%, and $\mathtt{division}$ increased by 22.2\%, and similar improvements are also found in \tapex.


\section{Preliminary}

\begin{figure*}[t]
\centering
    \includegraphics[width=0.95\linewidth]{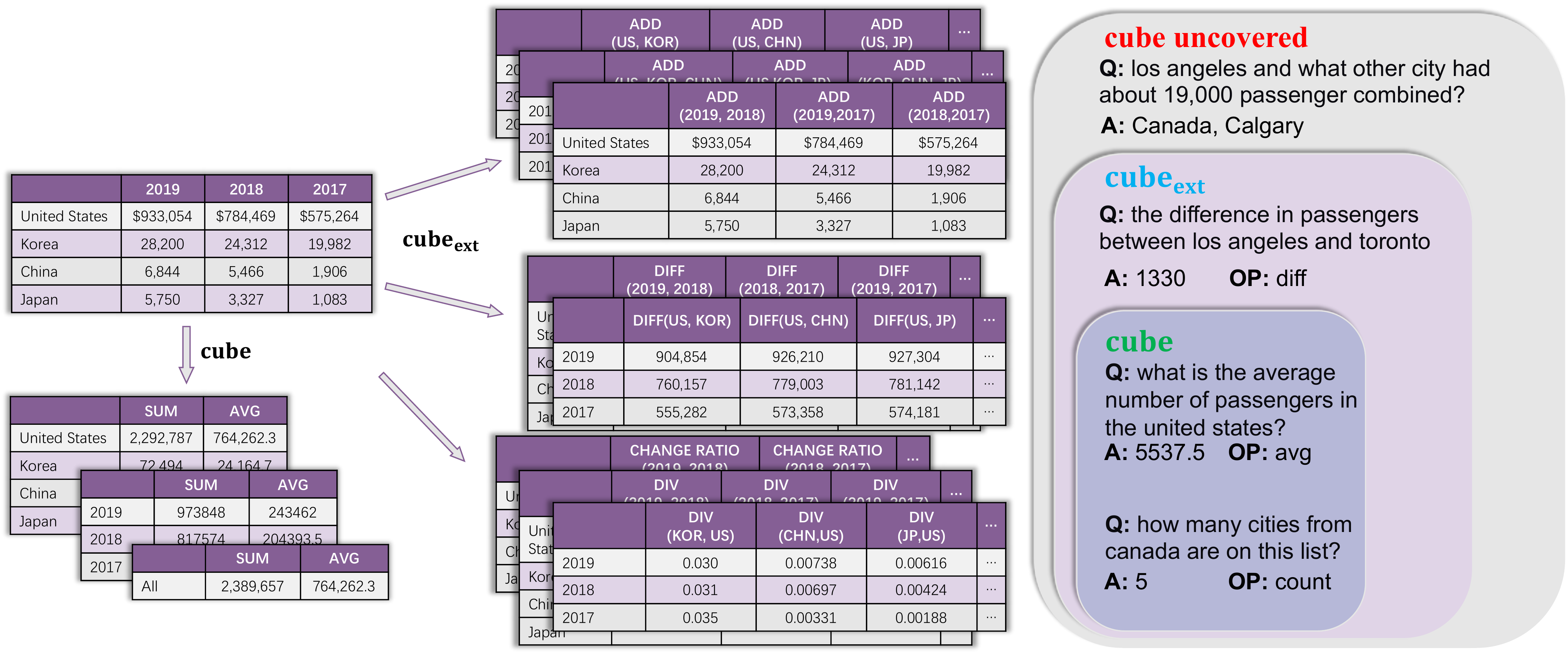}
\caption{Illustrtion of relationship between $\mathtt{cube}$, $\mathtt{cube_{ext}}$, and cube uncovered cases.}
\label{fig:cube_illustration}
\vspace{-0.32cm}
\end{figure*}


\subsection{Data Cube As Pre-computation}\label{subsec:data-cube-intro}
The concept of data cube is proposed in data mining, which can provides fast access to pre-computed, summarized data and thus benefit the analysis process~\cite{han2011data}. A data cube, which defined by dimensions and facts, allows stored data records to be modeled and viewed in multiple dimension. Each dimension corresponds to an attribute or a set of attributes, usually the perspectives or entities with respect to which an organization want to keep records. Each fact is numeric measure, organized by fact tables containing the name of the fact, usually storing value of some aggregate measure such as $\mathtt{count}$ and $\mathtt{sum}$. 

Take the cross table in Figure~\ref{fig:intro} for example. The numeric value stored in the table represents the annual accounts for a coorperation, the top rows which serve as the column header, the the first column serves as the row header. There are two dimensions in total: time dimension which records the yearly accounts, account type dimension which records multiple types of accounts.


Data cube inherently support operations including drill-down and roll-up, which allow the user to view the data at differing degrees. Also, the roll-up operation can be used to perform controllable data summarization along specified dimension(s). 

Semi-structured tables also share part of the features of stored records to generate a data cube. For example, the numeric values stored in a table cell can be treated as the stored records, while the header names or textual cell values can be treated as different dimensions corresponding to the numeric value records. 
The data cube naturally reflects some of the factual numerical for numerical reasoning over a given table, and can potentially helps tackling numerical reasoning problems over tabular data.

\subsection{Datasets}

\textbf{TAT-QA} dataset~\cite{zhu2021tat} includes in total 16,552 questions with hybrid context of both tables and textual content from real-world financial reports. Over 40\%(7341 of 16552) of the questions require numerical-reasoning skills including addition, substraction, multiplication, division, count, comparison, and compositional operations. TAT-QA also concludes the correct scale together with the string or numerical value to make up the final answer, which makes a unique challenge compared to other existing datasets.

\noindent \textbf{\wtq}~\cite{pasupat2015compositional} includes in total 22,033 questions focusing on table-only question answering. \wtq does not have annotations for numerical reasoning operations. The author manually picked 200 examples and classified them based on the types of operations required to answer the question. The case study reflects that the dataset contains considerable cases of complicated reasoning. The \textbf{Squall} dataset~\cite{shi2020squall} manually annotates SQL query on approximately 80\% of the cases in \wtq, which transfer the unsupervised task into a supervised semantic parsing task. The SQL annotations show that \wtq contains multiple types of operation, e.g., $\mathtt{count}$, $\mathtt{sum}$, $\mathtt{max}$, $\mathtt{min}$, $\mathtt{group}$, $\mathtt{abs}$, $\mathtt{avg}$.

Apart from the above two datasets studied in this paper, numerous tabular datasets are also closely related to numerical reasoning. Fin-QA~\cite{chen2021finqa} collects tables and context from financial domain, and raises financial-related questions including complex numerical reasoning. MultiHiertt~\cite{zhao2022multihiertt} and HiTAB~\cite{cheng2021hitab} include numerical reasoning examples on hierarchical table question answering. For table fact verification tasks~\cite{chen2019tabfact, aly2021feverous}, the numerical reasoning skills are also required for verifying whether the given statement about the table is entailed or refuted. 

However, the performance on existing methods tackling such tasks are still far from human performance, which indicates the numerical reasoning problems as an emerging challenge in tabular data.

\subsection{PLMs for Table QA}

PLMs have been widely leveraged in dealing with table QA tasks.
Recent works also prove that powerful PLMs with decoder can be well applied to table-text joint representation. On a well-recognized table QA benchmark \wtq, \uniskg~\cite{xie2022unifiedskg} achieves state-of-the-art denotation accuracy of 49.3\% among methods without extra pre-training; \tapex~\cite{liu2021tapex} with pre-training on SQL corpus surpasses the previous methods with a large margin, obtaining denotation accuracy of 57.5\%.


Represented as a form of pre-computed data cube, \ours can be easily integrated at the input phase and is theoretically suitable for most of the PLMs. Due to the shortcomings discussed in the Section~\ref{sec:intro}, in this paper, we mainly choose the PLMs with decoders and apply the generator based method in follow-up experiments over table QA tasks. The generator based method can effectively leverage the weakly-supervised data, and flexibly organized the cell value to generate the answer accordingly, e.g., to produce a piece of text inside a cell, or to produce an numeric value with its unit.


\noindent \textbf{Model Input/Output} For semi-structured tables, \tapex~\cite{liu2021tapex} designs a table linearization to feed the flatten sequence into the model directly. A flattened sequence is denoted as 
\begin{equation}
\begin{aligned}
    T^*= & \texttt{\small [HEAD]} ~|~ c_1 ~|~ {\cdots} ~|~ c_N \\
        & \texttt{\small [ROW]} 1 ~|~ r_1 ~|~ \texttt{\small[ROW]} 2 ~|~ r_2 ~|~ {\cdots} ~|~ r_M
\end{aligned}
\end{equation}
Notably, \tapex use a vertical bar $``|"$ to separate headers or cells in different columns. The final input will concatenate the textual context and the table sequence, and will then be fed into the model. The expected output of \tapex is the concatenation of answer(s) separated by comma which is generated autoregressively by the model. The table QA benchmarks in \tapex paper does not contain a more complex answer. But it is simple to conclude other forms of answer, e.g., using $``2~|~\rm{million}"$ to represent the answers with scale in TAT-QA.
\tapex proved to be effective in finding the factual supporting cells.

Our \ours takes the above model input and output setting. The pre-computed cube will also be flattened as a sequence, which will be discussed in Section~\ref{sec:rule-based}.

\section{\ours}
In this section, we first discuss how pre-computed data cubes help answering numerical reasoning questions over tables. Secondly, we point out the plain implementation via brute force are not favored due to explosive increasing cube size and time consumption. 
Following such observation, We present our approach for cube generation, and cube candidate selection. We will show that the proposed rule-based cube generation can cover a large portion of numerical reasoning required cases and remains efficient in search space at the same time. 
Further, in order to reduce the cube size when feeding to the PLMs, we adopt two ranking methods to perform a more effective filtering operation to pick out the most likely candidate cube results.


\subsection{Data Cube Application in Table QA}\label{subsec:cube_apply}


Following definition of cube, to apply data cube in Table QA, we need to decide dimension, and arithmetic/aggregation operator types.
Semi-structure tables contain rich information not only in cell values but also structural arrangement through headers or hierarchy~\cite{dong2022table,wang2021tuta}. 
For example, cells under the same header, or cells in the same table row are most likely to represent same type of information. 

Thus, a data cube for the table can be sets of aggregations over cells under same \textbf{column headers} or \textbf{row headers}. 
We denote such aggregation results over table headers as $\mathtt{cube}$, and they includes aggregation operations , such as $\mathtt{sum}$, $\mathtt{count}$, and $\mathtt{average}$. 

Nevertheless, such pre-computed results can't cover answers requiring non-aggregation operations, e.g., question asking the difference($\mathtt{diff}$) of two cell values or asking the summation result($\mathtt{add}$) of certain cells with specific filtering condition. In fact, the proportion of non-aggregation operations is even higher in some of the numerical reasoning datasets~\cite{zhu2021tat, zhao2022multihiertt, chen2021finqa}.
Based on such observation, We find it necessary to extend the operation types for pre-computed results, making it cover more operation types. 

We denote this extended pre-computed results as $\mathtt{cube_{ext}}$. The extended operators contain 2-operand operator, such as $\mathtt{add}$, $\mathtt{substraction}$, $\mathtt{division}$, $\mathtt{change~ratio}$. Notably, $\mathtt{add}$ and $\mathtt{sum}$ are in different groups of operators.
We use $\mathtt{sum}$ to represent an aggregation operator which sums up all the numeric value under one or many dimensions, and use $\mathtt{add}$ to represent an extended operator which adds up candidate numeric values after filtering.
As Figure~\ref{fig:cube_illustration} shows, in this paper, we only extend to \textbf{first-order} cube, i.e., we do not further perform operation over the generated cube items, which may include more compositional computation cases. We may exploit such practice in the future work.
We also include some computing pattern templates, e.g., $\mathtt{same~row}$ pattern which indicates the operand are all in the same row.
The extended operator types and the computing patterns for \ours is listed in Table~\ref{tab:rule-based}.

We will show $\mathtt{cube_{ext}}$ can cover a significant portion of the numerical reasoning cases in the following sections.

\begin{table}[t]
    \begin{subtable}{.5\textwidth}
    \small
    \centering
    \scalebox{0.9}{%
    \begin{tabular}{lll}
        \toprule[1.2pt]
        Operator        & aggr / ext       & Calculation       \\
        \hline
        count           & aggr             & $|{\rm{cell_{selected}}}|$                  \\
        \midrule
        sum             & aggr             & $\sum {\rm{cell_{selected}}}$ \\
        \midrule
        average         & aggr             & $1 / |{\rm{cell_{selected}}}| \cdot \sum {\rm{cell_{selected}}} $               \\ 
        \midrule
        add             & ext              & $\rm{cell_1 + \cdots + cell_k}$ \\
        \midrule
        diff            & ext              & ${\rm{cell_1 - cell_2}}$ \\
        \midrule
        div             & ext              & ${\rm{cell_1 / cell_2}}$ \\
        \midrule
        change ratio    & ext              & ${\rm{(cell_1 - cell_2) / cell_2}}$          \\
        \bottomrule[1.2pt]
    \end{tabular}%
    }
    \subcaption{Operator types for pre-computed cube}
    \end{subtable}
    \begin{subtable}{.5\textwidth}
    \small
    \centering
    \scalebox{0.8}{
    \begin{tabular}{ll}
    \toprule[1.2pt]
    Pattern     & \multicolumn{1}{c}{Operand}                                                \\
    \hline
    Same column & cells under one candidate column and all candidate rows \\
    \midrule
    Same row    & cells in one candidate row and all candidate columns \\
    \midrule
    All row     & all cells under one candidate column and all rows           \\
    \midrule
    All column  & all cells in one candidate row and all columns            \\
    \midrule
    Top-k row   & top-k rows' cells under one candidate column         \\
    \bottomrule[1.2pt]
    \end{tabular}%
    }
    \subcaption{Generation rules}
    \label{tab:rule-based-b}
    \end{subtable}
\caption{production rules}
\vspace{-0.3cm}
\label{tab:rule-based}
\end{table}

\subsection{Cube Generation}\label{sec:rule-based}
We aim to provide pre-computed results in model input phase so that the model can generate answers based on table cells as well as augmented cube information, bridging the gap in numerical reasoning. 
Also, to reduce the cube size, the cube generation is question-sensitive, i.e., for a given question and table pair, we generate one corresponding cube where cube items are most related to the question. For future work, we may try question-insensitive cube generation if the length of the cube is not a burden to the sequence length.

\noindent \textbf{Brute Force Generation} It is straightforward to use brute force method to traverse the whole search space and generate all the candidate results for the $\mathtt{cube_{ext}}$ of a given table. 
If the operation type and computing pattern to generate the expected answer is included in Table~\ref{tab:rule-based}, then the correct answer must exist in the generated $\mathtt{cube_{ext}}$. 

But brute force lead to two problems in the generation phase and model training phase. 
First, the explosive time consumption and space consumption for a large table. Suppose a table with $m$ rows and $n$ columns is given and we understand little about its structure, the theoretical total search space will be up to $O(n \cdot 2^m + m \cdot 2^n)$ for an aggregation operation, and $O(n \cdot m^2 + m \cdot n^2)$ for an extended 2-operand operation. 
Even if we rigorously follow the definition to generate the cube, we can still see from the Figure~\ref{fig:cube_illustration} that the size of the cube will easily exceed the size of the table itself. This is only the situation where the table is a simple matrix table which contains only two dimensions(time dimension, and country dimension), five rows and four columns.
For a multi-dimension table with much more columns and rows, we can imagine the time consumption and the search space will be intolerable.
Secondly, too many candidates can be a burden on the model input length and make it difficult for the model to leverage.

\noindent \textbf{Rule Based Generation} Due to the high expense and potential problems of brute force generation, it is worthy exploring the possibility to efficiently generate a pre-computed results for the Question-Table pair. We want the cube to be streamline so it can save input sequence's length. Meanwhile, it should cover most of the common numerical reasoning cases in the tasks. 


Following such observation, we design the generation rule for our pre-computing cube. We force the extracted candidate pre-computation to be same-row or same column operation. 
Also, we require the generated cube should be question-sensitive, i.e. to generate specific cube items closely related to the question. For instance, given a question answering $``$what is the difference in passengers between Los Angles and Toronto", then the cube should better not to conclude items with irrelevant operators, such as $\mathtt{count}$ and $\mathtt{sum}$.
We decide the candidate operators, headers, cells based on aligned mention in the question. 
We conclude the template for computing patterns of our rule-based generation in Table~\ref{tab:rule-based-b}. The template for computing patterns also reduces the search space of generation. 

The rule-based generation process can be concluded in three steps in general:
\vspace{-0.3cm}
\begin{itemize}
    \item[(i)] Decide operator type(s) by textual mention in questions.
    \item[(ii)] Find candidate columns and candidate rows by matching column headers, row header, and cell values with textual mention in questions.
    \item[(iii)] Try all the combinations using (1)~operator; (2) $k_{row}$ candidate rows and $k_{column}$ candidate columns; (3) computing patterns. Each computation result serves an item in the cube.
\end{itemize}

Such heuristic methods significantly shrink the time complexity to polynomial while sacrificing the coverage within a tolerable range. If a pre-computed cube for a given question and table pair contains one item that generate the correct numeric value to the answer, we treat it as a correct cube. The coverage is the correct cubes' proportion of total extracted cube numbers. 
(i) On TAT-QA dev set, we achieves coverage of approximately 70\% over all cases involving aggregation/arithmetic operation; 
(ii) On \wtq, we achieve coverage of 68\% over all the cube-extracted cases on the dev set and 62\% on the train set.



\noindent \textbf{Cube Serialization} The pre-computed and ranked cube will be flattened as a sequence in the end and this leads to designing cube serialization. As shown in Figure~\ref{fig:intro}, we present one item in the cube with its operator, the row header, the column header, the selected cell value, and the pre-computed answer. We design a naive cube linearization similar to the table linearization. And a flattened cube sequence is denoted as $C^*$.

\begin{equation}
\begin{aligned}
    C^*= & \texttt{\small [CUBE]}, \texttt{\small OPERATOR}, \\
         &   \texttt{\small {CH$_1$}}, {\cdots}, \texttt{\small {CH$_{kc}$}}, \texttt{\small {RH$_1$}}, {\cdots}, \texttt{\small {RH$_{kr}$}},  \\
         & \texttt{\small op{$_1$}}, \texttt{\small op{$_2$}}, \cdots, \texttt{\small op{$_m$}}, 
            \texttt{\small [ANSWER]} :  \texttt{\small answer}\\
\end{aligned}
\end{equation}
Here \texttt{\small [CUBE]} and \texttt{\small [ANSWER]} are special tokens indicating the start of the \ours and the answer; 
\texttt{\small {CH}}, \texttt{\small {RH}}, and \texttt{\small {op}} are abbreviations for \texttt{\small column header}, \texttt{\small row header}, and \texttt{\small operand};
$kc$, $kr$ are numbers of set of the headers of all the operands. 

Notably, in Table~\ref{tab:rule-based} we concluded all the computing patterns for a pre-computed result. So all the operands in one \ours must share either the same column headers or same row headers.

\subsection{Cube Ranking}\label{sec:neural ranker}

Although the extracted pre-computed cube results' size is within polynomial level, it still can be non-trivial and causes the same dilemma as the brute force generation. 
To overcome the problem, we propose two methods for ranking the cube results, including a heuristic method and a neural-based method. 


\noindent \textbf{Heuristic Ranking} We use the proposed cube linearization method to generate cube sequences and calculate the text similarity~\cite{ramos2003using} between the question and the flattened cube sequences. The ranks of pre-computed results are decided by the similarity, i.e., the result with a higher similarity will have a higher rank.

\noindent \textbf{Neural-based Ranking} The neural-based ranking process is modeled as a binary classification problem following table fact verification tasks. Given the concatenation of a question, a pre-computed cube result, and a table, the neural-based model needs to give its prediction of whether the pre-computed result is the correct answer. We employ a BART-base architecture and use the similar configuration of \tapex's experiments on TabFact~\cite{chen2019tabfact}. We re-rank the candidates using both the label and the output logits.

\begin{figure}[t]
    \begin{center}
    \includegraphics[width=1.2\linewidth]{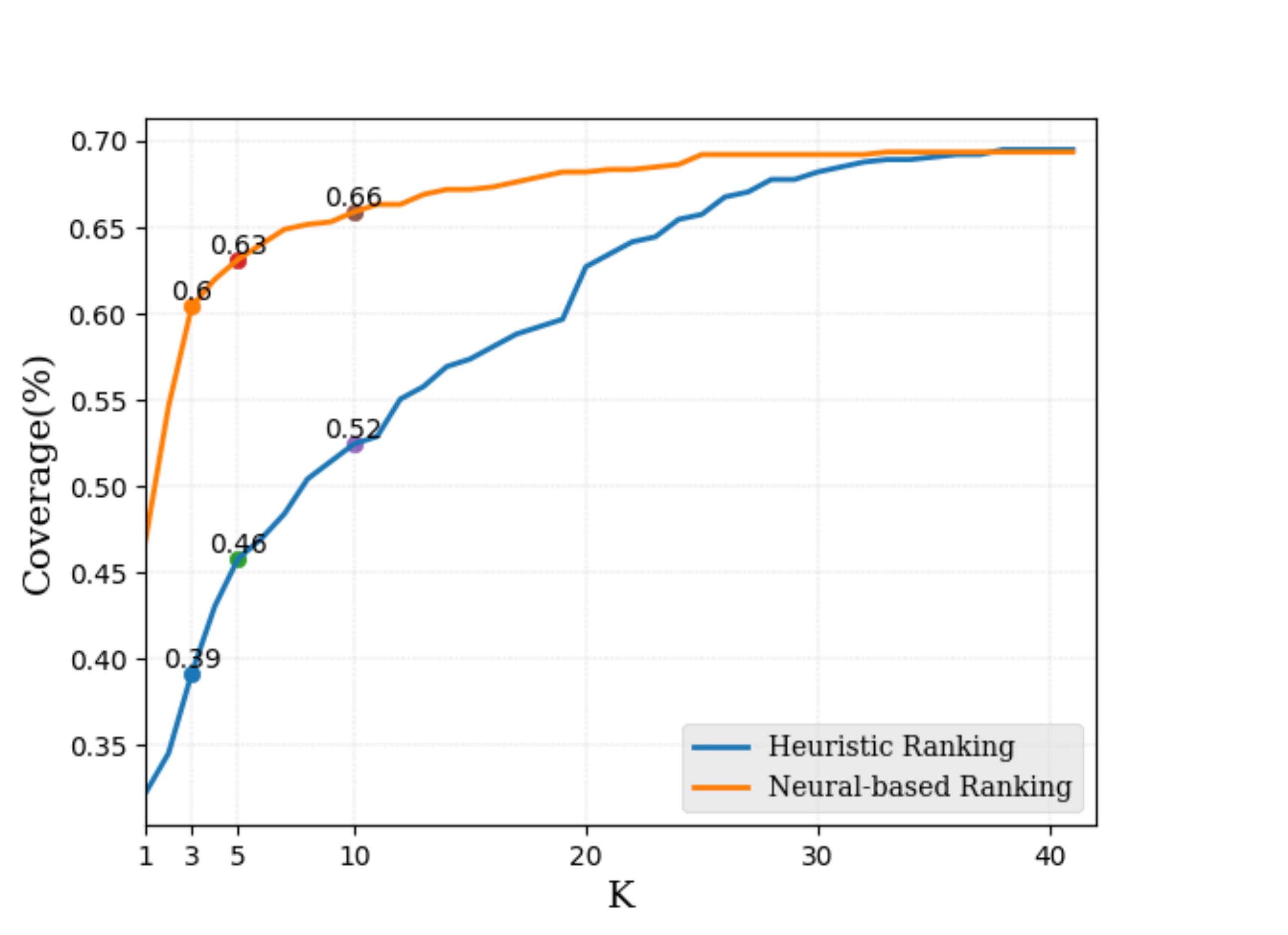}
    \end{center}
    \vspace{-0.3cm}
\centering
\caption{Coverage on validation data of tat-qa, where \textbf{K} stands for picking top k pre-computed results as input.}
\label{coverage on tat}
\vspace{-0.32cm}
\end{figure}

Both heuristic ranking and neural-based ranking method uses $\mathtt{cube_{ext}}$, and we denote the top $k$ picked pre-computed cube items as $\mathtt{cube_{k}}$. 




\subsection{Coverage Discussion}
In section~\ref{sec:rule-based}, we propose a rule-based cube generation method. And in section ~\ref{sec:neural ranker}, we further introduce two scoring methods for evaluating cube confidence and re-ranking the cube. To prove that our method covers most of the reasoning cases, we test the coverage of extracted cubes on validation data of TAT-QA\cite{zhu2021tat}.

As is shown in Fig \ref{coverage on tat}, increasing the number of pre-computed cube results will result in a greater coverage, eventually reaching about 70\%. Additionally, with the use of neural ranker, the coverage under the same number of pre-computed cube results has increased significantly.

We also make an analysis on the failed cases in 100 random samples of \wtq dev test. 
We observe that the upper bound of cube generation does not outperform current rule-based cube generation method by a large margin: only 15\% of the fail-to-cover cubes are caused by current rules.
For more details,  please check Appendix~\ref{sec:failed to extract}.


\section{Experimental Results and Analysis}
In this section, we describe the details of \ours and evaluate the effectiveness of \ours on two table-related questions answering benchmarks. 
\subsection{Experimental Setup}
\noindent \textbf{Datasets} We evaluate on TAT-QA\cite{zhu2021tat}, and \wtq \cite{pasupat2015compositional}. 
TAT-QA requires multiple numerical reasoning skills, such as addition, subtraction, multiplication, division, and compositional operations. TAT-QA contains in total of 16552 questions with hybrid context of both tables and textual content from real-world financial reports. 
\wtq is one of the most used table QA benchmarks focusing on table-only question answering tasks. It also requires reasoning capabilities, including aggregate, superlative, and arithmetic operations. 
On TAT-QA, We apply the exact match and F1 score used in TAT-QA paper. On \wtq, we apply denotation accuracy as the evaluation metric.

\noindent \textbf{Baselines} We summarize the baseline methods in short below. 
(i) On TAT-QA, we include two different families of methods for comparison: current SOTA Roberta-based method TagOP\cite{zhu2021tat}, and sequence-to-sequence generator-based method including BART\cite{lewis2020bart}, and \tapex \cite{liu2021tapex}.
(ii) on \wtq, SOTA model \tapex on multiple table qa and table fact verification tasks and BART\cite{lewis2020bart}.

\noindent \textbf{Serialization} For each baseline generator-based model, we adopt the same serialization in \tapex, which concatenate the input questions and the linearized table. For extracted and ranked cube items, we use cube serialization discussed in Section~\ref{sec:rule-based}. Like \tapex, we also separate different cubes using a vertical bar $``|"$.

\noindent \textbf{Implementation Details} For \wtq, the fine-tuning is set up to 50 epochs with initial learning rate as $3\times10^-5$. The effective batch size is 128. For TAT-QA, the fine-tuning is set up to 50 epochs with initial learning rate as $5\times10^-5$. The effective batch size is 24. For pre-computed cube item number $k$, we choose $k=10$ for TAT-QA, and $k=5$ for \wtq.

\begin{table}[t]
\centering
\scalebox{0.85}{
\begin{tabular}{p{3.5cm}p{1.2cm}<{\centering}p{1.2cm}<{\centering}}
    \toprule[1.2pt]
    Models     & EM         & F1 \\
        \midrule
        TagOP      & 55.2       & 62.7\\
        \midrule
        BART-Large & 38.8       & 46.7 \\
        \hspace{0.2cm} $w.$ \ours + HR & 55.2       & 63.7 \\
        \hspace{0.2cm} $w.$ \ours + NR & \textbf{57.1}       & \textbf{65.6} \\
        \midrule
        Tapex-Large      & 41.5       & 49.6 \\ 
        \hspace{0.2cm} $w.$ \ours + HR      & 56.9       & 65.8\\ 
        \hspace{0.2cm} $w.$ \ours + NR & \textbf{57.7}       & \textbf{66.2}\\
    \bottomrule[1.2pt]
\end{tabular}%
}
\caption{Exact match and F1 scores on TAT-QA dev set.}
\label{tab:tatqa-main-results}

\end{table}

\begin{table}[t]
\centering
\scalebox{0.85}{
\begin{tabular}{p{3.5cm}p{1.2cm}<{\centering}p{1.2cm}<{\centering}}
    \toprule[1.2pt]
        Models                             & \multicolumn{1}{c}{Dev} & \multicolumn{1}{c}{Test} \\
            \midrule
            BART-Large                         & 37.2               & 38.0                \\
            \hspace{0.2cm} $w.$ \ours + HR               & 42.1              & 40.0                \\
            \hspace{0.2cm} $w.$ \ours + NR               & \textbf{42.9}             & \textbf{40.3}                \\
            \midrule
            Tapex-Large*                         & 58.9              & 57.5              \\
            \hspace{0.2cm} $w.$ \ours + HR               & \textbf{59.7}               & \textbf{59.6}                \\
            \hspace{0.2cm} $w.$ \ours + NR      & 59.3               & 59.2                \\
    \bottomrule[1.2pt]
\end{tabular}%
}
\caption{Denotation accuracies on \wtq dev set and test set. We reimplement the \tapex on \wtq{} and surpass the reported results on the dev set in \tapex paper(57.0\% $\rightarrow$ 58.9\%).}
\label{tab:wtq-main-results}

\end{table}

\subsection{Main Results}

Table \ref{tab:tatqa-main-results}, Table \ref{tab:wtq-main-results} present the evaluation results of various models' performance, where \textbf{HR} and \textbf{NR} are abbreviations for \textbf{H}euristic \textbf{R}anking and \textbf{N}eural-based \textbf{R}anking.
Across all instances, we observe the marginal increase in performance. 
(i) On the dev set of TAT-QA, \ours registers the EM of 57.7\% and the F1 score of 66.2\%, surpassing the baseline \tapex by 16.6\% and TagOP by 3.5\%. 
(ii) On both dev and test set of \wtq, \ours also achieves new state-of-the-art denotation accuracy of 59.7\%(+0.8\%) and 59.6\%(+2.1\%).

On each benchmark, we also study \ours's effect over arithmetic/aggregation cases. 
For TAT-QA, we test over all the annotated arithmetic cases and extract the operator following TagOP's practice. Moreover, on TAT-QA, F1 score is always equal to the exact match on arithmetic/aggregation involved cases, thus we only report EM in the table. 
For \wtq, the distribution among arithmetic cases and aggregation cases is extremely unbalanced. To study \ours's effect on each operator, we separate out a subset of the original dataset using annotations in the Squall dataset. Squall~\cite{shi2020squall} manually annotates SQL query corresponding to the question and table, covering 80\% examples for train and dev set of the \wtq{}. We determine the operator by the aggregation/arithmetic keywords appearing in the SQL query and get the performance over each operator. 

The results are presented in Table~\ref{tab:wtq-op-results} and Table~\ref{tab:tatqa-operator-results}. The model with \ours substantially outperform on each type of the operator. Again, \tapex shows its power in aggregation operator: among extracted $\mathtt{count}$ operations, \tapex without \ours achieves denotation accuracy of 64.0\%, which outperforms the BART baseline by over 30\%.

Further, because the proportion and the absolute number of \wtq{} samples requiring summation, average, difference and other operations are quite low, it is not conducive for \ours's training. Despite such negative factors, using \ours still bring performance boost on each operator. On TAT-QA which contains more arithmetic cases, \ours{} outperforms on every operator by a large margin.

\begin{table}[t]
\centering
\scalebox{0.80}{
\begin{tabular}{p{2.8cm}p{0.55cm}p{0.55cm}p{0.55cm}p{0.55cm}p{0.55cm}p{0.55cm}}
\toprule[1.2pt]
Model & CR. & Avg & Sum & Diff & Div & Arith. \\
\toprule
BART-Large     & 0.0              & 2.8         & 1.8                & 4.3            & 5.6          & 2.6            \\
\hspace{0.1cm} $w.$ \ours + HR     & 69.7             & \textbf{62.4}        & 26.3               & 38.7           & 5.6          & 44.0           \\
\hspace{0.1cm} $w.$ \ours + NR     & \textbf{78.7}             & 55.3        & \textbf{40.4}               & \textbf{40.9}           & \textbf{27.8}         & \textbf{47.5}           \\
\midrule
Tapex-Large     & 0.6              & 3.5         & 1.8                & 6.0            & 5.6          & 3.5            \\
\hspace{0.1cm} $w.$ \ours + HR    & 78.1             & 53.2        & \textbf{33.3}               & \textbf{51.9}           & \textbf{33.3}         & \textbf{50.0}           \\
\hspace{0.1cm} $w.$ \ours + NR     & \textbf{79.4}             & \textbf{53.9}        & \textbf{33.3}               & 48.5           & 27.8         & 49.6	\\
\bottomrule[1.2pt]
\end{tabular}%
}
\caption{Exact match on arithmetic/aggregation involved cases of TAT-QA dev set. \textbf{Arith.} represents all cases involving arithmetic operations. \textbf{CR.} stands for the \textbf{c}hange-\textbf{r}atio operator.}
\label{tab:tatqa-operator-results}
\end{table}

\begin{table}[t]
\centering
\scalebox{0.8}{
\begin{tabular}{llllcc}
    \toprule[1.2pt]
        Models & Diff & Sum & Avg & Count & Total\\
            \midrule
            BART-Large                            & 2.4           & 2.5     & 10.0 & 33.6 & 30.2           \\
            \hspace{0.2cm} $w.$ \ours + HR    & 21.4           & \textbf{20.0}      & \textbf{40.0} & \textbf{46.1}    & \textbf{43.5}      \\
            \hspace{0.2cm} $w.$ \ours + NR   & \textbf{33.3}           & 15.0      & \textbf{40.0}  & 45.7   & \textbf{43.5}      \\
            \midrule
            Tapex-Large                           & 19.0           & 10.0     & 10.0 & 64.0      & 58.4      \\
            \hspace{0.2cm} $w.$ \ours + HR                 & \textbf{33.3}           & \textbf{22.5}      & \textbf{20.0} & \textbf{65.4}    & \textbf{61.0}    \\
            \hspace{0.2cm} $w.$ \ours + NR      & \textbf{33.3}    & 17.5           & 10.0      & 65.3         &  60.6 \\
    \bottomrule[1.2pt]
\end{tabular}%
}
\caption{Performance on different operators of \wtq dev set. The number of cases involving difference, summation, average and count is 41, 40, 10 and 709 respectively.}
\label{tab:wtq-op-results}

\end{table}






\section{Related Work}
\noindent \textbf{Table QA with PLMs} There are a variety of works applying PLMs on Table QA. To perform question answering over semi-structured tabular data, prior work formulated the problem into a semantic parsing task, and adopted semantic parsers to operate over tables~\cite{yin2020tabert, shi2018incsql, liang2018mapo, cheng2021fortap}. 
To better encode and represent the tabular data, work also focused on: (i) table input featurization, which may conclude extra information about tabulat data, e.g., column/row embeddings based on cell location~\cite{wang2021tuta, herzig2020tapas, eisenschlos2021mate}; (ii) structural-aware encoding, which designs visualization matrix for structure-pruned attention of transformers~\cite{wang2021tuta, eisenschlos2021mate} or produce embedding in a row-wise/column-wise manner~\cite{yin2020tabert}; (iii) table pre-training using collected tabular data corpus~\cite{yin2020tabert, liu2021tapex, herzig2020tapas, eisenschlos2021mate}.
Recent work has shown the potential of directly using PLMs for table QA. Classified by the model architecture, the encoder-based methods~\cite{herzig2020tapas, eisenschlos2021mate} applies multiple classification head on the encoding representation and generate the answer. The generator based methods expect the decoder to autoregressively generate the ground truth answer without extra design~\cite{liu2021tapex, xie2022unifiedskg}. We mainly focus on the possbility to improve the performance on the generator based method, especially its numerical reasoning skills.

\noindent \textbf{Numerical Reasoning over tabular data} Numerical reasoning is important in different NL tasks~\cite{dua2019drop, zhu2021tat, zhao2022multihiertt}. In table domain, tables usually contain numeric values which are well-organized. Moreover, in several end-user tools, the spreadsheet formulas in cells imply the numerical relationships among the table cells~\cite{dong2022table}. Therefore, a wide range of tasks require numerical reasoning over tabular data, such as table-to-text~\cite{suadaa2021towards, cheng2021hitab}, formula prediction~\cite{cheng2021fortap, chen2021spreadsheetcoder} and table fact verification~\cite{chen2019tabfact, aly2021feverous} and table question answering~\cite{pasupat2015compositional}. 
Recently, various benchmarks are proposed to solve table QA problems and containing a large proportion of numerical reasoning examples~\cite{chen2021finqa, zhu2021tat, zhao2022multihiertt}.

\noindent \textbf{Retrieval based ODQA} Most existing methods deals with open domain QA(ODQA) by retrieving evidence over documents~\cite{chen2017reading}, knowledge graph triples~\cite{verga2021adaptable} and collection of question-answer pairs~\cite{chen2022augmenting, xiao2021open} to aggregate the answer using retrieved evidence. Different from these works, our method do not directly transform the table QA into a retrieval problem; instead, the model need to leverage the pre-computed cube and may perform post-processing during answer decoding, such as adding a scale string, and further comparison based on the cube items.



\section{Conclusion}
In this paper, we focus on numerical reasoning problems over tabular question answering. We propose our method, namely \ours, which automatically extracted pre-computed results for a given question and a table following the designed cube generation rules. 
The pre-computed cube is serialized and fed to the model in the input phase, mitigating the gap in numerical reasoning skills for PLMs.
\ours is tested over multiple table QA benchmarks using a encoder-decoder architecture, and achieves new SOTA performance on each of them. Further analysis shows the performance boost mainly comes from the numerical reasoning examples in the benchmarks. In the future, we may exploit a more complex cube which can cover more numerical reasoning cases, e.g. a second-order cube, which has not yet been included in this paper.

\bibliographystyle{acl_natbib}
\bibliography{acl_natbib}
        
\clearpage
\appendix

\section{Failed-to-extract cases}\label{sec:failed to extract}

We categorize cases of failed extraction into four categories:
(i) \textbf{Outside Knowledge}: We need to have both the knowledge provided by the form as well as external knowledge in order to answer such questions. Take nt-7969 as an instance, the question answering the times of an athlete compete in the game. While in the original table(which is shown in Figure~\ref{fig:outside_knowledge}), one row is denoteed as $``$DNF" which means $``$do not finish". Such cases need outside knowledge in sports terminology and are hard to generate the answer.
(ii) \textbf{Non-number Pattern}: The answers to these questions contain non-number patterns, while the pre-computed cube results are limited to numerical data only.
(iii) \textbf{Rule-uncovered Cases}: As stated in Section~\ref{subsec:cube_apply}, current designed rule for cube generation only consider naive first-order data cubes. Thus, the extracted cubes can not include compositional computation. Moreover, for unusual numeric formats, such dates, length in ft., or magnitude, it is non-trivial to design a general rule and currently such cases are not covered.
(iv) \textbf{Other Cases}: Answers are either incorrectly annotated or the reasoning process is unclear.

\vspace{-0.32cm}
\subsection{Outside Knowledge}

\textbf{Question:} how many times did imma clopes compete?

\begin{figure}[t]
    \begin{center}
    \includegraphics[width=3.in]{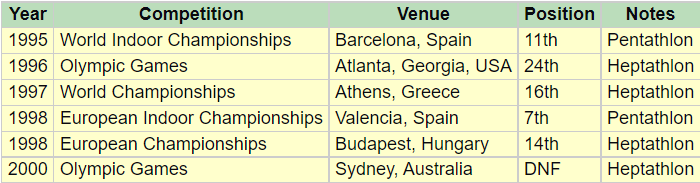}
    \end{center}
    \vspace{-0.3cm}
\caption{outside knowledge}
\label{fig:outside_knowledge}
\vspace{-0.32cm}
\end{figure}

\noindent
\textbf{Table:} See Figure \ref{fig:outside_knowledge}

\noindent
\textbf{Answer:} 5

\noindent
\textbf{Analysis:} The question answering the times of an athlete compete in the game. While in the original table(which is shown in Figure~\ref{fig:outside_knowledge}), one row is denoteed as $``$DNF" which means $``$do not finish". Such cases need outside knowledge in sports terminology and are hard to generate the answer.
\vspace{-0.32cm}
\subsection{Non-number Pattern}

\textbf{Question:} what is the difference between the time air uganda commenced operations and skyjet airlines commenced operations?

\begin{figure}[t!]
    \begin{center}
    \includegraphics[width=3.in]{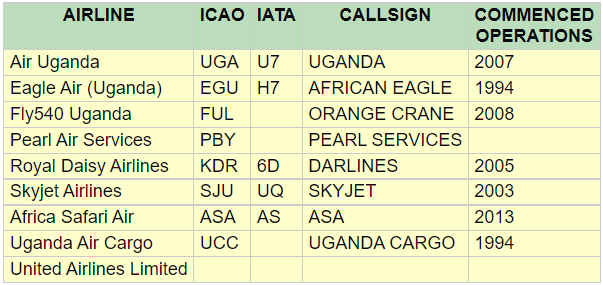}
    \end{center}
    \vspace{-0.3cm}
\caption{non-number pattern}
\label{fig:nonnumber_pattern}
\vspace{-0.32cm}
\end{figure}

\noindent
\textbf{Table:} See Figure \ref{fig:nonnumber_pattern}

\noindent
\textbf{Answer:} 4 years

\noindent
\textbf{Analysis:} The answer to the question includes a non-number pattern. Thus, the value "4" is included in our precomputed cube results, however, the answer "4 years" is not. 





\begin{table}[t]
\centering
\scalebox{0.75}{
\begin{tabular}{l|l}
\hline
\multirow{4}{*}{\makecell[l]{Outside\\Knowledge~\\(9\%)\\~\\}}
& e.g., nt-7969 \\
& \textbf{question}: how many times \\
& did imma clopes compete?\\
& \textbf{answer}: 5\\
\hline
\multirow{6}{*}{ \makecell[l]{Non-number\\Pattern~\\(1\%)\\~\\~\\~\\}}
& e.g., nt-6701 \\
& \textbf{question}: what is the difference \\
& between the time air uganda \\
& commenced operations and \\
& skyjet airlines commenced operations?        \\
& \textbf{answer}: 4 years\\
\hline
\multirow{4}{*}{ \makecell[l]{Rule-uncovered\\Case~\\(15\%)\\~\\}}
& e.g., nt-2401 \\
&\textbf{question}: what was the average time \\
& for the americans? \\
& \textbf{answer}: 4:19:41\\
\hline
\multirow{4}{*}{ \makecell[l]{Other \\Case~\\(3\%)\\~\\}}
& e.g. nt-10206 \\
& \textbf{question}: how many years ago \\
& did ne-yo play as mixx? \\
& \textbf{answer}: 8\\
\hline
Correct Case~(72\%)                     & \multicolumn{1}{c}{-}    \\
\hline
\end{tabular}%
}
\caption{Detailed analysis on \ours fail-to-cover cases. We randomly pick 100 samples in \wtq dev set and manually check the extraction results.}
\label{tab:error-case}
\end{table}
\vspace{-0.32cm}
\begin{figure}[t!]
    \begin{center}
    \includegraphics[width=3.in]{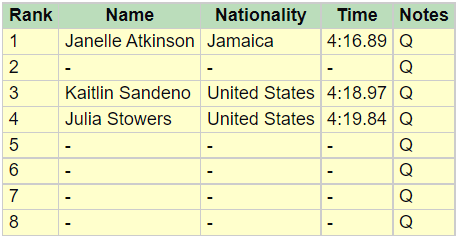}
    \end{center}
    \vspace{-0.3cm}
\caption{rule-uncovered case}
\label{fig:ruleuncovered_case}
\vspace{-0.4cm}
\end{figure}
\begin{figure}[t!]
    \begin{center}
    \includegraphics[width=3.in]{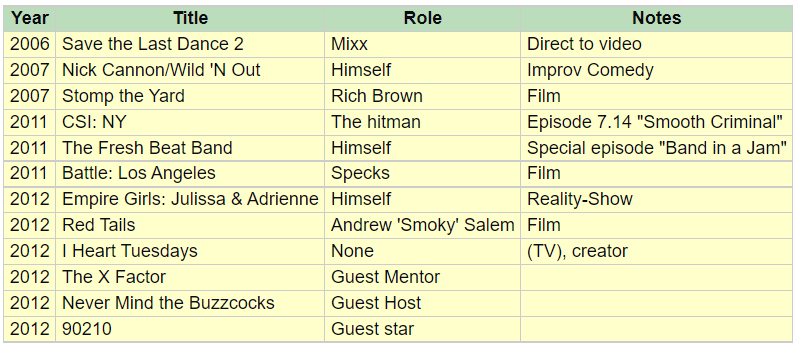}
    \end{center}
    \vspace{-0.32cm}
\caption{other case}
\label{fig:other_case}
\end{figure}

\subsection{Rule-uncovered Case}

\textbf{Question:} what was the average time for the americans?

\noindent \textbf{Table:} See Figure \ref{fig:ruleuncovered_case}

\noindent \textbf{Answer:} 4:19.41

\noindent
\textbf{Analysis:} Due to the fact that the data format in the table is a date, the calculation of the result requires parsing complex data.
\subsection{Other Case}
\textbf{Question:} how many years ago did ne-yo play as mixx?
\noindent
\textbf{Table:} See Figure \ref{fig:other_case}
\noindent
\textbf{Answer:} 8

\noindent \textbf{Analysis:} The question answering $``$how many years ago" and the answer is 8, which implies such sample is annotated in year 2014. However, it is too hard to know such information, making reasoning almost impossible.

\end{document}